\documentclass[aps,prl,preprint,showpacs,superscriptaddress,amsmath]{revtex4-1}
\usepackage{amsfonts}
\usepackage{mathtools}
\usepackage{amsthm}
\usepackage{amsmath,scalerel}
\usepackage{bm}
\usepackage{siunitx}
\usepackage{braket}
\usepackage{color}
\usepackage{float}

\newcommand{\mbm}[1]{\bm{{\rm #1}}}

\begin{document}
\title{Functionalized high-speed magnon-polaritons resulting from the magnetic antenna effect}
\author{Kenta Kato}
\affiliation{Department of Material Engineering Science,
Graduate School of Engineering Science, Osaka University,
1-3 Machikaneyama, Toyonaka, Osaka 560-8531, Japan}
\author{Tomohiro Yokoyama}
\email[E-mail me at: ]{tomohiro.yokoyama@mp.es.osaka-u.ac.jp}
\affiliation{Department of Material Engineering Science,
Graduate School of Engineering Science, Osaka University,
1-3 Machikaneyama, Toyonaka, Osaka 560-8531, Japan}
\author{Hajime Ishihara}
\email[E-mail me at: ]{ishi@mp.es.osaka-u.ac.jp}
\affiliation{Department of Material Engineering Science,
Graduate School of Engineering Science, Osaka University,
1-3 Machikaneyama, Toyonaka, Osaka 560-8531, Japan}
\affiliation{Department of Physics and Electronics,
Graduate School of Engineering, Osaka Prefecture University,
1-1 Gakuen-cho, Naka-ku, Sakai, Osaka 599-8531, Japan}
\affiliation{Center for Quantum Information and Quantum Biology, Osaka University,
1-3 Machikaneyama, Toyonaka, Osaka 560-8531, Japan}

\date{\today}

\begin{abstract}
Magnon-polaritons (MPs) refer to a light--magnon coupled state and can potentially act as information carriers,
possibly enabling charge-free computation.
However, the light--magnon coupling is inherently weak.
To achieve sufficiently strong coupling, a large ferromagnet or coupling with a microwave cavity is necessary.
Herein, we theoretically propose a fundamental platform for magnonic and magnon--optical information storage devices
and discuss the transport properties of MP's.
The proposed multi-layered structure overcomes the aforementioned issues.
Owing to the waveguide modes, magnons placed in a nanometer-thin layer are strongly coupled with light,
exhibiting rich functionalities of thick-layer MPs via the `magnetic antenna effect'.
Thus, the thin-layer MPs are faster, and the direction is switchable.
The results of this study will enable the integration of ferromagnetic micro and nanostructures for
MP-based information devices without any restrictions due to cavities.
\end{abstract}
\maketitle

In magnetic materials, elementary excitation arises from collective motion of spin precession;
this phenomenon, which is equivalent to a quasiparticle, is called a spin wave or magnon.
The magnon is an potentially viable candidate for low-energy-consumption devices
because it can carry~\cite{Kajiwara10,Chumak14,Chumak15,Oyanagi19}, process~\cite{Chumak14,Tabuchi15,Lachance-Quirion19},
and store~\cite{XZhang15} the information of spin without charge carrier transport.
The spin current from magnons is proportional to the density of excited magnons $\rho_{\rm m}$ and
the group velocity of the magnons $v_{\rm g} = (1/\hbar) (\partial \varepsilon /\partial k)$.
Thus, there exist two significant issues affecting the efficiency of magnonic devices:
(i) the increase in magnon current with increasing density and velocity, and (ii) switchable control over the current direction.
However, typical magnon densities are significantly lower than the typical electron density in metals.
The typical velocity of a magnon is also considerably lower than the Fermi velocity of electrons.
To overcome this issue, recent studies have focused on the engineering of magnon velocity~\cite{CLiu18,CLiu19} and
transport~\cite{Onose10,Matsumoto14,Murakami17}.
One method to increase the velocity of magnons is to form magnon-polaritons (MPs) via strong coupling between
magnons and microwaves~\cite{YCao15,Tabuchi14,XZhang14}.
The spin-light coupling is significantly weaker than the coupling between an electric dipole and light.
Recent studies have achieved ultra-strong coupling for MPs using optical cavities,
and significant splitting of the dispersion relation of magnetic microspheres has been reported~\cite{Tabuchi14,XZhang14}.
Such cavity-based MPs in magnetic microspheres exhibit rich properties, such as bistability~\cite{YPWsng18,Hyde18},
coherent perfect absorption~\cite{DZhang17,XZhang19}, level attraction~\cite{Harder18,PCXu19,WYu19,Rao19NJP},
the microwave Hall effect~\cite{Rao19NC}, and information communication between
magnetic segments in cavities~\cite{Lambert16,Bai17,Rameshti18,ZZhang19,JLi19}.
However, to ensure information transport via magnons, it is necessary to achieve strongly coupled MPs,
which we refer to as `strong MPs', in one- or two-dimensional integrable magnetic materials.

\begin{figure}[H]
\begin{center}
\includegraphics[width=150mm]{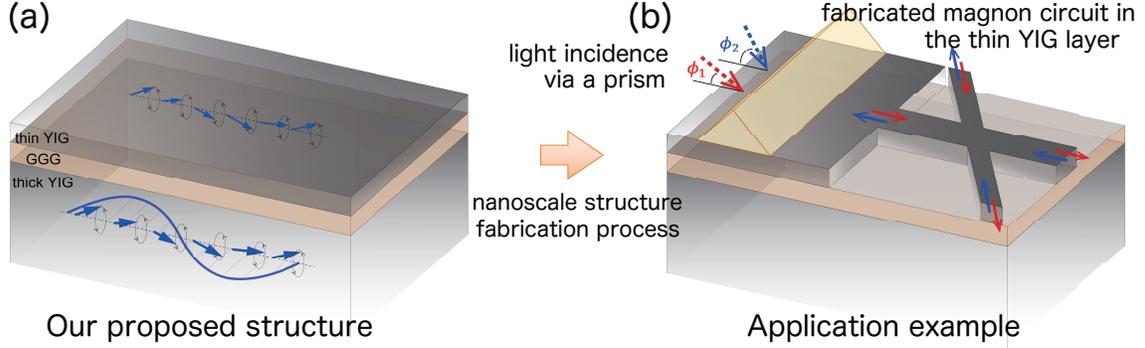}
\caption{
Schematic of the proposed multi-layer structure for magnonic devices.
(a) Proposed structure comprising thin yttrium iron garnet (YIG), thin gadolinium gallium garnet (GGG), and thick YIG layers.
The thick YIG layer acts as an antenna to capture photons via the formed waveguide mode.
The photons are coupled with the magnons in the thin YIG layer via the magnon--magnon interaction.
Thus, the thin-layer magnons can form polaritons.
(b) Imaginary processed structure. It is expected that even when using a thin YIG layer to fabricate circuits,
the MPs in the thin YIG layer will demonstrate high-speed and controllable magnon transport in circuits,
without any leakage of the magnon current. This is due to the functionality of MPs.
(Refer to the text immediately above the Summary section.)
To form polaritons using the waveguide mode, a microwave with a prism or fiber coupling is employed.
As an example of circuit fabrication, an X-shaped junction is presented.
}
\label{fig:model}
\end{center}
\end{figure}

Several systems have been studied with the aim of discovering new applications for strong MPs using cavities.
To the best of our knowledge, however, contemporary studies have not exploited the potential of the interplay between
the spatial degrees of freedom of the magnons and the radiation field.
To incorporate this ingredient, we propose a multi-layered two-dimensional magnetic film,
where the spatial profiles of the magnon wave and radiation field and their interplay are considered key components.
Subsequently, a nonlocal treatment for the microwave response from magnons is formulated.
This approach, which considers the nonlocal response of magnons to the radiation field, provides a new application prospect for strong MPs
and is expected to offer several degrees of freedom in the spatial design of sample structures.
Our formulation of a nonlocal theory is based on the quantum microscopic perspective of magnons.
(This scheme can be extended to nonlinear magnon response in a straightforward manner.) 
The spatial interplay between magnons and the field was designed under both dielectric and magnetic environments.
Thus, a multi-layered structure of magnetic and dielectric materials is proposed, as shown in Fig.\ \ref{fig:model}.
The structure comprises thin ($\sim$ 100 nm) and thick ($\sim 1 \mathrm{mm}$) magnetic layers.
A dielectric layer with an appropriate thickness separates the magnon wavefunctions in the two magnetic layers.
Figure \ref{fig:model}(a) depicts the model considered in the present calculations, while (b) shows an imaginary processed structure.
This structure is based on three important concepts.
First, the thin magnetic layer can be processed to fabricate integrable magnonic circuits.
Second, the `waveguide modes' formed by the interfaces significantly enhance the strength of the MPs in the magnetic layers.
Third, an interlayer magnon--magnon interaction results in the `antenna effect',
which transfers the rich functionalities of the thick-layer MPs (such as the switching function of the transport direction) to the thin-layer MPs.
Importantly, the separation of magnons by the dielectric layer is essential for the antenna effect and magnon circuit fabrication,
as discussed in subsequent sections.
Moreover, the cavity-less structure offers significant advantages for further development,
such as MP network fabrication on a substrate and MP control using a small magnet tip.
The formulated nonlocal theory (refer to the Methods section)
for magnons indicates that a spatial correlation between the magnons and waveguide modes
results in strong MPs (as discussed in the following sections).
A combination of these aspects enables the realization of highly efficient and switchable solid-state devices,
without the requirement for charge carrier transport.

\section{Results and Discussion}
\begin{figure}[H]
\begin{center}
\includegraphics[width=160mm]{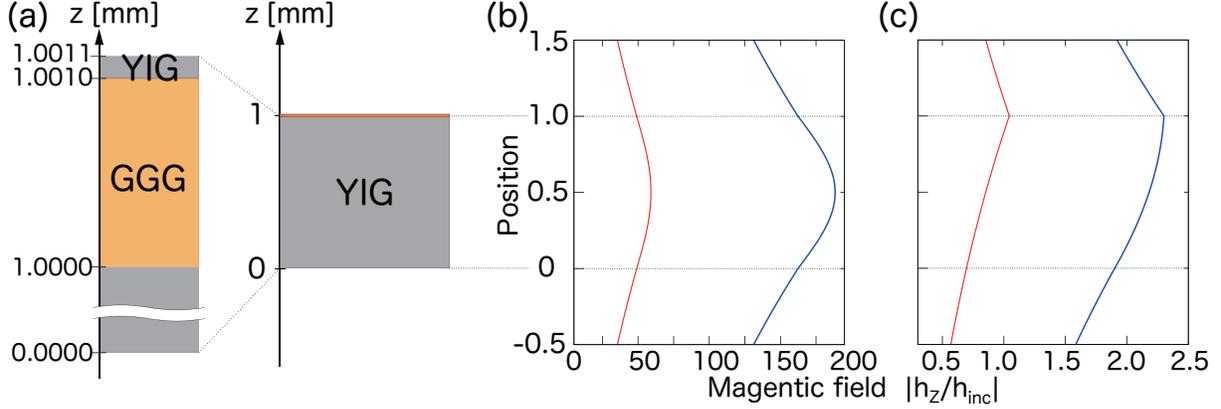}
\caption{
(a) Schematic of the cross-section of the proposed multi-layer structure with
$100\, \mathrm{nm}$ YIG/$1\, \mathrm{\mu m}$ GGG/$1\, \mathrm{mm}$ YIG layered structure.
(b,c) Spatial profiles of the transverse (blue) and longitudinal magnetic fields (red lines) originating from
(b) the thick and (c) thin-layer magnons.
The field amplitude is normalized by the incident field $h_{\rm inc}$.
Other parameters are as follows: $\mu_0 M_{\rm s} = 175\, \mathrm{mT}$, $\lambda_{\rm ex} = 1.76 \times 10^{-8}\, \mathrm{m}$,
$\epsilon_{\rm M} = 15$, $\epsilon_{\rm b} = 12$. $H_0^{(1)} = H_{\rm M}$, and $H_0^{(2)} = 1.5H_{\rm M}$.
The unit of the magnetic field is $H_{\rm M} \equiv \hbar \omega_{\rm M}/(g\mu_{\rm B}\mu_0) = M_{\rm s}$.
}
\label{fig:field}
\end{center}
\end{figure}

For the proposed multi-layered structure shown in Fig.\ \ref{fig:model}(a),
we assume $100\, \mathrm{nm}$ and $1\, \mathrm{mm}$ yttrium iron garnet (YIG) layers.
These two layers are separated by a $1\, \mathrm{\mu m}$ gadolinium gallium garnet (GGG) layer,
which is the paramagnetic insulator and standard material for heterostructures with YIG.
A schematic of this structure is presented in Fig.\ \ref{fig:field}(a).
Thick YIG films are favorable for stronger MP coupling, which results in larger magnon currents. 
However, to ensure suitable fabrication quality, it is important to verify that
even a considerably thinner (nanometer-order) YIG film can function suitably under the antenna effect of the thick layer.
Thus, we select a thickness of 100 nm, which can be easily achieved and processed using standard fabrication techniques.

The saturation magnetization is $\mu_0 M_{\rm s} \simeq 175\, \mathrm{mT}$ for YIG~\cite{Daniel},
where the frequency corresponding to magnetization is
$\omega_{\rm M} = g \mu_{\rm B} \mu_0 M_{\rm s} /\hbar \simeq 4.89 \times 2\pi \, \mathrm{GHz}$.
$\mu_{\rm B}$ is the Bohr magneton, the g-factor is $g=2$,
and the coherence length of the exchange interaction is $\lambda_{\rm ex} \simeq 1.76 \times 10^{-8} \mathrm{m}$.
The dielectric constant is $\epsilon_{\rm M} = 15$ for YIG~\cite{Sadhana} and $\epsilon_{\rm b} = 12$ for GGG~\cite{Lal77}.
The magnon is excited in each YIG layer, and the interlayer magnon--magnon interaction is attributed to the dipole--dipole interaction.
An incident microwave $h_{\rm inc}$ is introduced onto the surface of the thick layer.
The microwave is applied to the multi-layered structure
even for an in-plane wavenumber $k_{\parallel}$ larger than the wavenumber $q_0 = \omega/c$ in vacuum.
This can be realized by utilizing an evanescent prism mode on the structure, or a fiber coupling at the edge surface.
For the calculation of the fields inside and outside the sample, we place a field inside the thick YIG layer and
consider the boundary conditions of the microwave through Green's function to
determine the self-consistent fields in other layers (refer to the Methods section).
We apply an in-plane gradient magnetic field with $H_0^{(1)} = H_{\rm M}$ and $H_0^{(2)} = 1.5 H_{\rm M}$ to the thin and thick layers,
respectively, to obtain the level (anti-)crossing of thin-layer magnons and thick-layer MPs.
Here, $H_{\rm M} \equiv \hbar \omega_{\rm M}/(g \mu_{\rm B} \mu_0)$.
Such a spatial gradient of the magnetic field can be realized using micromagnets or current-induced fields.
(The use of different magnetic materials for the two layers also affords similar level crossings.)

Here, we emphasize the role of the GGG layer.
The GGG spacer layer enables engineering dispersion.
Even for a system without a GGG layer, namely a single $100\, \mathrm{nm} + 1\, \mathrm{mm}$ YIG film, circuit fabrication is possible.
However, the magnon wavefunctions spread along the longitudinal direction.
Hence, the spatial gradient of the applied magnetic field, which is indispensable for manipulating dispersion crossing, becomes ineffective.
In addition, the magnon becomes insensitive to the fabricated circuit,
because the hundred-nanometer-scale modulation of thickness is not significant for the magnons in the millimeter-scale YIG film.
Therefore, the GGG layer plays a critical role in the functioning of the proposed system.
It disconnects the magnon wavefunction in the thin layer from that in the thick layer,
and a thickness of 1 {$\mu$}m is sufficient for this purpose.
The decay length of the spin dipole--dipole interaction for the magnon wavenumber under consideration is
significantly greater than the aforementioned thickness.
Hence, a slight, micrometer-order modulation of the thickness for the GGG layer does not affect the antenna effect significantly.

In the proposed structure, the `waveguide modes' and the `antenna effect' impart the thin-layer MP transport with functionalities
that are essential for practical applications.
We consider the waveguide mode in a simple YIG layer with thickness $d$.
The two waveguide modes, i.e., the transverse electric (TE) and magnetic (TM) modes, are formed within
$q < k_{\parallel} < q_0$ in the YIG layer, which decays exponentially in vacuum, where $q = q_0/ \sqrt{\epsilon_{\rm M}}$.
The waveguide modes satisfy the condition $k_z d + {\rm Arg} (r_{\xi}) = n\pi$ for the $\xi = \text{TE, TM}$ modes.
Here, $k_z = \sqrt{q^2 - k_{\parallel}}$.
The reflection coefficient $r_{\xi}$ at the interfaces depends on the wavenumber $(k_{\parallel},k_z)$.
Here, $n$ denotes the index of the waveguide mode, and the modes require millimeter-scale thickness, $d \simeq \pi/k_z \sim \mathrm{mm}$.
Although this condition is slightly different for multi-layer structures, the behavior of the mode remains qualitatively unaffected.

In the proposed structure, the magnons in the thick and thin YIG layers induce both a longitudinal magnetic field and
a transverse electromagnetic field (Figs.\ \ref{fig:field}(b) and (c)).
The longitudinal field results in an interlayer magnon--magnon interaction.
The fields from the thick-layer magnons are almost symmetric in the $z$-direction, whereas
the transverse field from the thin-layer magnons exhibits non-symmetric behavior owing to the difference of dielectric constants.
Hence, considering the nonlocal response of magnons is essential.
We formulate a general nonlocal response theory for magnons~\cite{kinoshita19,book:Ishihara},
which considers a self-consistent relation between Maxwell's equations and the constitutive equation with
$\bm{E}$, $\bm{H}$, $\bm{M}$, $\bm{P}$, and nonlocal susceptibility (see Methods section).

\begin{figure}[H]
\begin{center}
\includegraphics[width=145mm]{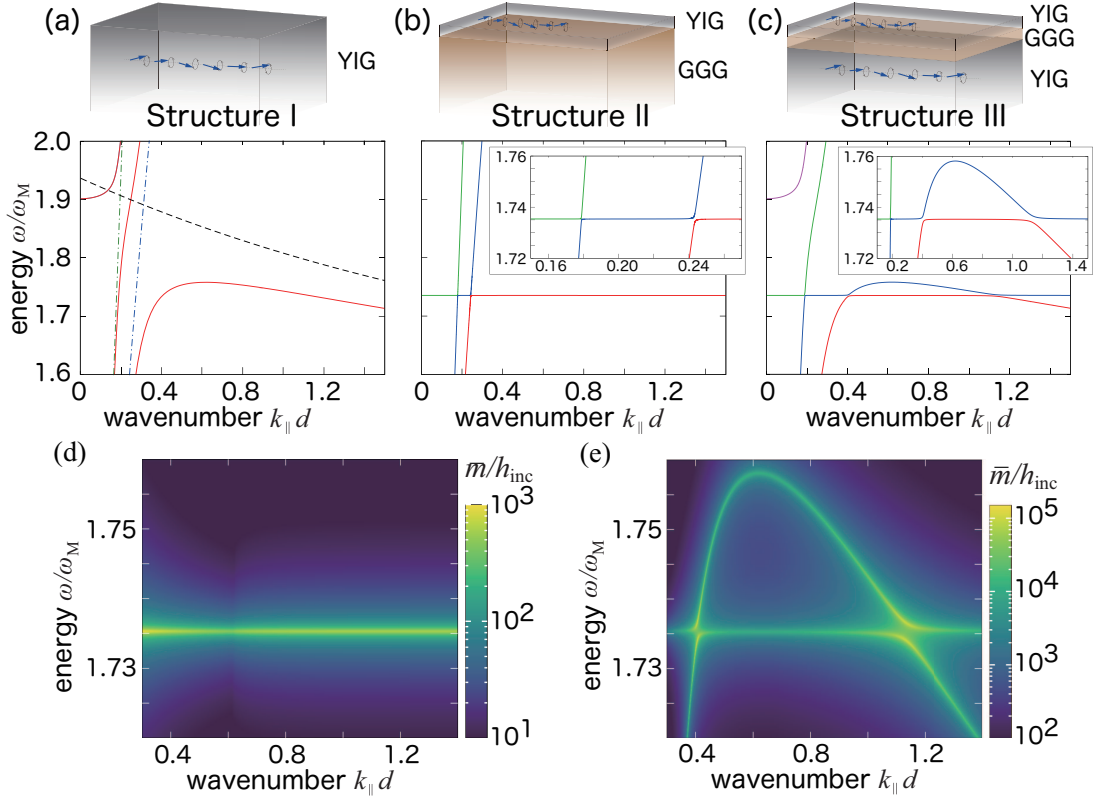}
\caption{
(a,b,c) Dispersion relations $\omega (k_{\parallel})$ of MPs in the single $1 \mathrm{mm}$ YIG film (a, Structure I);
$100\, \mathrm{nm}$ YIG/$1\, \mathrm{mm}$ GGG structure (b, Structure II); and
proposed structure, $100\, \mathrm{nm}$ YIG/$1\, \mu \mathrm{m}$ GGG/$1\, \mathrm{mm}$ YIG (c, Structure III).
The corresponding structures are shown above the panels. 
The wavenumber is normalized by the thickness of the system, $d = 1\, \mathrm{mm}$.
The YIG layers exhibit in-plane saturation magnetization.
The other parameters are the same as those in Fig.\ \ref{fig:field}.
The external static magnetic fields are applied along the in-plane magnetization.
The solid lines indicate the MPs in panels (a), (b), and (c).
In (a), the black dotted line indicates the bare magnon, and
the blue and green dotted lines represent the TE and TM modes, respectively.
In (b) and (c), the colors of the solid lines are clearly distinguishable.
The insets of (b) and (c) present zoomed views.
(d,e) Magnetization amplitude $\bar{m}$ deviated from the saturation magnetization of the thin YIG in
Structure II (d) and III (e), respectively.
The amplitude is normalized by the incident magnetic field $h_{\rm inc}$.
The magnetization $\bar{m}$ is enhanced on the dispersion of MPs.
}
\label{fig:dispersion}
\end{center}
\end{figure}

Engineering the dispersion relation enables control over magnon propagation.
To demonstrate this, we consider three types of layered systems with
YIG: Structure I: $1\mathrm{mm}$ YIG film, Structure II: $100\mathrm{nm}$ YIG/$1\mathrm{mm}$ GGG, and
Structure III: $100\mathrm{nm}$ YIG/$1\mathrm{\mu m}$ GGG/$1\mathrm{mm}$ YIG (the proposed structure).
The YIG layers exhibit in-plane magnetization.
The antenna effect originates from the dispersion of the thick YIG layer,
and the waveguide modes require the system to possess a certain thickness.
Figures \ref{fig:dispersion}(a) and (b) present the dispersion relations of the MPs in Structures I and II, respectively.
In Structure I, the bare magnon exhibits a negative slope~\cite{Kalinikos}.
The thick film spontaneously forms the TE and TM waveguide modes, which results in strong MPs and three dispersions.
We refer to these as the thick-layer upper, middle, and lower MPs (denoted as thick-UMP, thick-MMP, and thick-LMP, respectively).
At $0.1 <k_{\parallel} d<0.5$, the group velocity of the three MPs is significantly higher than that of the bare magnon.
The in-plane wavenumber $k_{\parallel}$ can be tuned by adjusting the incident angle.
With an increase in the incident angle, the propagation direction of the thick-LMP changes around $k_{\parallel}d = 0.6$.
In the proposed structure (Structure III), the functionalities of the thick-layer MPs are transferred to
the thin-layer magnons via the antenna effect.
By contrast, in Structure II, the dispersion exhibits direct coupling between the thin-layer magnons and the TE/TM modes.
Their strengths are $\Delta_{\rm TE} \simeq 3.3 \times 10^{-3} \omega_{\rm M}$ and
$\Delta_{\rm TM} \simeq 1.7 \times 10^{-3} \omega_{\rm M}$, respectively.
If the $100\, \mathrm{nm}$ YIG layer is isolated from the GGG layer, where the waveguide mode is not formed,
the coupling strength between the magnon and photon is negligible.
In this case, the dispersion of the bare magnon should be almost flat.

The properties of the thin-layer MPs in the proposed structure are provided by the waveguide modes (transverse field) and
the antenna effect, based on the interlayer magnon--magnon interaction (longitudinal field).
Figure \ref{fig:dispersion}(c) shows the dispersions of the MPs in Structure III.
The thick YIG layer captures the microwaves and provides functionality to the thin-layer magnons via the antenna effect.
The functionalities imparted to the thin-layer magnons can be tuned using the external magnetic fields.
As shown in Fig. \ref{fig:dispersion}(c), the thin-layer magnons couple with the thick-MMP at $k_{\parallel}d \approx 0.2$ and
with the thick-LMP at $k_{\parallel}d \approx 0.4$ and $1.2$.
Interestingly, at the latter two anti-crossings, the MPs exhibit forward and backward propagation, respectively.

\begin{figure}[H]
\begin{center}
\includegraphics[width=130mm]{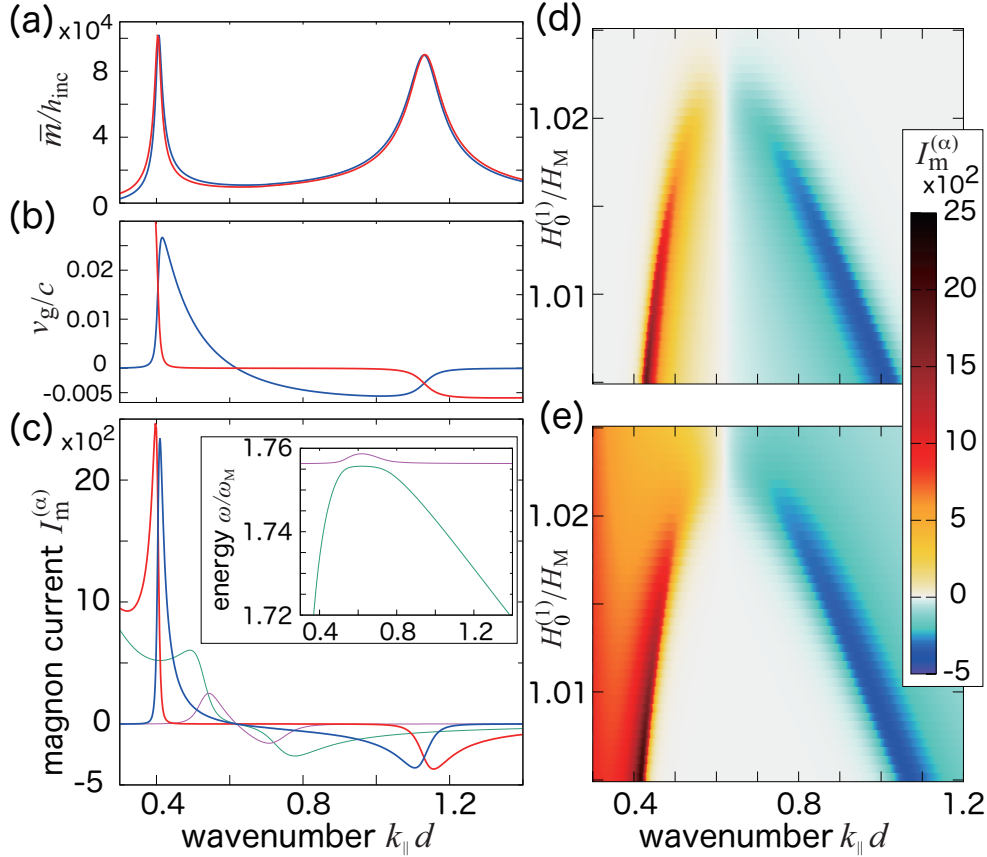}
\caption{
Magnon current in thin YIG layer of Structure III. 
(a) $\bar{m}$ in Fig.\ \ref{fig:dispersion}(e) along dispersion of two MPs in Fig.\ \ref{fig:dispersion}(c).
(b) Group velocity $v_{\rm g}$ of thin-layer MP dispersion, as shown in Fig.\ \ref{fig:dispersion}(c).
The velocity is normalized by the velocity of light in vacuum, $c$.
(c) (Normalized) magnon current $I_{\rm m}^{(\alpha)}$ from thin-layer MPs when external fields are $H_0^{(1)} = H_{\rm M}$ (thick lines) and $\approx 1.02 H_{\rm M}$ (thin lines).
The red and blue lines in (a), (b), and (c) denote
the lowest and second branches in Fig.\ \ref{fig:dispersion}(c), respectively.
The green and magenta lines and inset show $I_{\rm m}^{(\alpha)}$ and the dispersion relation
when $H_0^{(1)} \approx 1.02 H_{\rm M}$, respectively.
(d,e) The magnon currents, $I_{\rm m}$, from the second (d) and lowest (e) branches of thin-layer MPs in a plane of the external magnetic field are represented as functions of $H_0^{(1)}$ and the in-plane wavenumber, $k_{\parallel}d$.
\label{fig:current}
}
\end{center}
\end{figure}

We evaluate the magnon current in the proposed structure.
The group velocity is evaluated based on the dispersion, whereas the magnon density is evaluated based on
the magnetization amplitude of the deviation from the saturation magnetization, $\mbm{m} (\mbm{r}) = \mbm{M} (\mbm{r}) - \mbm{M}_{\rm s}$.
Figures \ref{fig:dispersion}(d) and (e) present the spatial integral of the induced magnetization in the thin YIG layer,
$\bar{m} = \sqrt{\frac{1}{S_0 d_0} \int_{d_0} dz \int_{S_0} dxdy |\mbm{m} (\mbm{r})|^2}$, 
for Structures II and III as functions of $\omega$ and $k_{\parallel}$, respectively.
The enhancement in $\bar{m}$ implies the generation of a large magnon density.
Here, $S_0$ and $d_0$ denote the area of the YIG unit cell and the thickness of the thin layer, respectively.
In Structure II, although the induced magnetization $\bar{m} (k_{\parallel}, \omega)$ follows the dispersion of the bare magnon,
the amplitude is not large.
However, in Structure III, $\bar{m}$ is considerably increased owing to the interlayer magnon--magnon interaction
appearing on the dispersions of the thin layer of the bare magnon and thick-LMP.
This large $\bar{m}$ results in a large magnon current in the thin YIG.
Along the dispersion $\omega^{(\alpha)} (k_\parallel)$ in Structure III,
we plot the wavenumber dependence of the induced magnetization $\bar{m}^{(\alpha)}$ and the group velocity $v_{\rm g}^{(\alpha)}$,
as shown in Figs.\ \ref{fig:current}(a) and (b), respectively.
$\alpha$ is an index of the MP branches $\omega^{(\alpha)} (k_{\parallel})$.
The magnon density in the thin YIG layer is enhanced owing to the strong coupling with the thick-LMP.
The MP velocity is increased up to $0.01$ times the light speed in YIG.
This velocity is in the order of $10^6 \mathrm{m/s}$, which is $10^3$ times
the typical (exchange) magnon velocity~\cite{HXia98,HYu12,Yamanoi13,CLiu18}.

Figure \ref{fig:current}(c) shows the magnon currents in the MPs, which are expressed as
\begin{equation}
I_{\rm m}^{(\alpha)} (k_{\parallel}) = v_{\rm g}^{(\alpha)} (k_{\parallel}) \times
\bar{m}^{(\alpha)} (k_{\parallel}, \omega^{(\alpha)}(k_{\parallel}) ).
\label{eq:magnoncurrent}
\end{equation}
Following the peaks of $\bar{m}^{(\alpha)}$ and the sign of $v_{\rm g}^{(\alpha)}$,
the magnon current in the thin YIG layer exhibits peak and dip behaviors at $k_{\parallel}d \approx 0.4$ and $1.2$, respectively.
For the applied static magnetic fields, $H_0^{(1)} = H_{\rm M}$ and $H_0^{(2)} = 1.5 H_{\rm M}$,
it is necessary to tune the incident angle of the microwave from $k_{\parallel}d \approx 0.4$ to $1.2$ to switch the magnon current direction. 
The distance between the peak and dip is rather large in this case.
However, this distance can be reduced by controlling the shifts in the thin-layer magnon energy.
To this end, two anti-crossing points can be modulated by tuning the applied field $H_0^{(1)}$ in the thin layer.
We examine a sweep of the thin-layer magnon energy by $H_0^{(1)}$ up to the dispersion edge of the thick-LMP.
Figs. \ref{fig:current}(d) and (e) demonstrate the magnon current $I_{\rm m}^{(\alpha)}$ from the two MP branches, respectively.
With a slight increase in $H_0^{(1)}$, the distance between the peaks and dips of $I_{\rm m}^{(\alpha)}$ can be reduced.
At $H_0^{(1)} > 1.02 H_{\rm M}$, the peaks and dips are unified, and the magnon current is reduced significantly.
The thin lines and the inset of Fig.\ \ref{fig:current}(c) indicate the magnon current and dispersion, respectively,
when $H_0^{(1)} \approx 1.02 H_{\rm M}$. Here, the peak and dip are sufficiently detectable, and
the sign can be altered by a slight shift in $k_{\parallel}$.
This condition can be deemed appropriate for magnon current switching.
When $H_0^{(1)}$ decreases, the distance between the peaks and dips increases.
For a large $k_\parallel d$, the magnon--magnon interaction is suppressed owing to
the exponential decay of the longitudinal field for short-wavelength microwaves.
Then, the dip disappears when $H_0^{(1)}$ is reduced.

The dispersion exhibits anti-crossing with the thick-MMP at $k_{\parallel}d \approx 0.2$ (see Fig.\ \ref{fig:dispersion}(c)).
The peak and dip of $I_{\rm m}^{(\alpha)}$ for the thick-LMP are broad,
whereas the peak of the thick-MMP is extremely sharp and isolated (outside the window in Fig.\ \ref{fig:current}(c)).
However, we do not consider this peak, because the characteristics of such a sharp peak are not useful for practical applications.
If the magnon circuit is fabricated in the thin layer, as shown in Fig.\ \ref{fig:model}(b),
the internal magnetic field and the magnetization are slightly varied.
The sharp peak is highly sensitive to such changes and is difficult to control.

For the processed structures illustrated in Fig.\ \ref{fig:model}(b), some parts of YIG are removed.
In such areas, the dispersion of MPs is almost equivalent to that shown in Fig.\ \ref{fig:dispersion}(a).
The thick-LMP may induce a large magnon current.
However, at the dispersion edge ($k_\parallel d \approx 0.65, \omega \approx 1.76 \omega_{\rm M}$),
the magnon current is suppressed due to the low velocity.
Thus, the leakage magnon current is expected to be reduced.
This is also an important advantage of the proposed structure.
Details regarding the behaviors of magnons in the processed thin layer cannot be obtained 
only from the present study; these behaviors should be elucidated based on corresponding models.
We are currently investigating this issue using numerical techniques based on nonlocal response theory.

\section{Conclusions}
The magnon is a promising candidate for low-energy-consumption devices.
To ensure effective device operation, the magnon current should be increased by
increasing the density and velocity and acquiring switchable control over the direction of the current.
One method to increase the velocity of the magnon is the formation of MPs.
However, the light--magnon coupling is inherently weak.
To address this bottleneck, we proposed a multi-layered magnetic structure to demonstrate
a large and switchable magnon current in the integrable thin magnetic layer.
We considered two crucial mechanisms: the `waveguide mode' and `antenna effect'.
The large magnon current and its switchable functionality are supported by these two phenomena.
The formation of the waveguide mode is attributed to the spatial structure of the dielectric constant,
which considerably enhances the light--magnon coupling caused by the spatial interplay between the magnons and microwaves.
The antenna effect, originating from the magnon--magnon interaction (longitudinal field), maps
the functionality of the thick-layer MPs to the nanoscale thin-layer magnons.
By analyzing the dispersion relations of the MPs in this system and
numerically calculating the magnon currents in the nanometer-scale thin film based on nonlocal response theory,
we demonstrated that the proposed multi-layered structure can successfully exhibit the desired features mentioned above.
The antenna effect arising from the thick-LMP produces a broad peak and dip in the magnon current
as a function of the incident angle.
This is desirable for circuit fabrication in the thin layer.
Our proposed approach thus presents the intriguing possibility of developing advanced magnonic technologies.

This work is supported in part by JSPS KAKENHI 16H06504 in Scientic Research on Innovative Areas Nano-Material Optical-Manipulation", 
JSPS KAKENHI 18K13484 and 18H01151.
T.Y. is supported by the Toyota Riken Scholar from the Toyota Physical and Chemical Research Institute. 

The authors declare no conflicts of interest.

\section{Methods}

\subsection{Magnon}
\label{sec:spinwave}
The spin wave in magnetic materials is described by the exchange interaction
$\hat{\mathcal{H}}_{\rm ex} = - \sum_{\langle i,j \rangle} J_{ij} \hat{\mbm{S}}_i \cdot \hat{\mbm{S}}_j$.
Here, we consider a uniform ferromagnetic film, $J_{ij} = J_0 >0$,
and apply the Holstein--Primakoff transformation~\cite{HP40},
$\hat{S}^z_i = S - \hat{b}_i^\dagger \hat{b}_i$,
$\hat{S}^-_i = \sqrt{2S} \hat{b}_i^\dagger (1-\frac{\hat{b}_i^\dagger \hat{b}_i}{2S})$,
$\hat{S}^+_i = \sqrt{2S} (1-\frac{\hat{b}_i^\dagger \hat{b}_i}{2S}) \hat{b}_i$. 
On removing the ground state energy of the ferromagnet,
we obtain the magnon Hamiltonian $\hat{\mathcal{H}}_{\rm{sw}}$, which can be rewritten in terms of the boson operators 
$\hat{b}_{\mbm{k}}$ and $\hat{b}_{\mbm{k}}^\dagger$ as
$\hat{\mathcal{H}}_{\rm ex} \simeq \hat{\mathcal{H}}_{\rm sw} - \delta N J S^2$ with
\begin{equation}
\hat{\mathcal{H}}_{\rm sw} = 2S \sum_{\mbm{k}}
\left[ \tilde{J} (0) - \tilde{J} ({\mbm{k}}) \right] \hat{b}_{\mbm{k}}^\dagger \hat{b}_{\mbm{k}}.
\label{eq:exchange}
\end{equation}
Here, $\hat{b}_{\mbm{k}} = \frac{1}{\sqrt{N}} \sum_i \hat{b}_i e^{-i\mbm{k} \cdot \mbm{R}_i}$.
$\tilde{J} ({\mbm{k}}) = \sum_{i^\prime} J_0 e^{-i \mbm{k} \cdot \mbm{R}_{i^\prime} }$ is
the summation of all neighboring sites $\mbm{R}_{i^\prime}$ of the localized electrons at $\mbm{R}_i$;
$\delta$ and $N$ represent the number of neighboring sites and the total number of sites, respectively.
$S$ is the size of the localized spin at each site.
This boson describes the quantum of the spin wave, i.e., the so-called magnon.
A constant term can be reduced without a loss of generality.
The exchange interaction originates from the interaction between the localized electrons,
which is a part of the longitudinal electric fields.
However, the dipole--dipole interaction is considered in Maxwell's equations.

The magnetic fields also contribute to the magnon under the Zeeman effect:
$\hat{\mathcal{H}}_{\rm Z} = -g\mu_{\rm B}\mu_0 \sum_{i} \hat{\mbm{S}}_i \cdot \mbm{H}(\mbm{r}_i;t)$,
where $\mu_{\rm B}$ is the Bohr magneton and $g$ is the $g$-factor.
The magnetic field $\mbm{H} (\mbm{r};t)$ comprises an external static field, $\mbm{H}_0$,
an incident microwave, $\mbm{h}_{\rm inc} (\mbm{r};t)$, and the response field, $\mbm{h}_{\rm res} (\mbm{r};t)$.
The response field is generated by the induced magnons.
We consider $\mbm{h} (\mbm{r};t) = \mbm{h}_{\rm inc} (\mbm{r};t) + \mbm{h}_{\rm res} (\mbm{r};t)$ and
define the non-perturbative and perturbative Hamiltonians as
$\hat{\mathcal{H}}_0 = \hat{\mathcal{H}}_{\rm sw} + g\mu_{\rm B}\mu_0|\mbm{H_0}|\sum_{\mbm{k}}\hat{b}_{\mbm{k}}^\dagger \hat{b}_{\mbm{k}} $ and
$\hat{\mathcal{H}}^\prime = -g\mu_{\rm B}\mu_0 \sum_{i} \hat{\mbm{S}}_i \cdot \mbm{h} (\mbm{r}_i;t)$, respectively.
For $\hat{\mathcal{H}}_0$, the energy dispersion of the magnons is 
\begin{equation}
E_{\mbm{k}} = 2S [\tilde{J} (0) - \tilde{J} ({\mbm{k}})] + g\mu_{\rm B}\mu_0 |\mbm{H}_0|
\simeq \hbar \omega_{\rm M} \lambda_{\rm ex}^2 k^2 + \hbar \omega_{\rm H},
\end{equation}
which is shifted by the external magnetic field $|\mbm{H}_0|$.
Here, $\lambda_{\rm ex} = \sqrt{ 2S |\nabla_k^2 J(k)|/ (\hbar \omega_M) }$ is the exchange length of the spin wave.
$\hbar \omega_{\rm M} = \hbar \gamma \mu_0 |\mbm{M}_{\rm s}|$ is the magnetization energy, where
$\gamma = g\mu_{\rm B}/\hbar$ is the gyromagnetic ratio and the saturation magnetization is $\mbm{M}_{\rm s}$.

\subsection{Nonlocal response of magnon}
\label{sec:nonlocal}
Based on the quantum description of magnons, we formulate the microscopic theory of magnon response.
The constitutive equation for magnetic response is 
$\mbm{m} (\mbm{r}_i;\omega) = \sum_j \bar{\chi} (\mbm{r}_i,\mbm{r}_j;\omega) \mbm{h} (\mbm{r}_j;\omega)$
with the nonlocal susceptibility~\cite{book:Cho}.
Here, $\mbm{m}$ is the deviation of the magnetization from the saturation magnetization, and
$\mbm{m} (\mbm{r};t) = \mbm{M} (\mbm{r};t) - \mbm{M}_{\rm s}$.
The susceptibility is obtained from the Hamiltonian $\hat{\mathcal{H}}_0 + \hat{\mathcal{H}}^\prime$, based on the linear response theory~\cite{Kubo57}:
\begin{equation}
\bar{\chi} (\mbm{r}_i,\mbm{r}_j;\omega)
= -\frac{\mu_0}{v} \sum_n \left[
  \frac{\mbm{\mu}_{0n} (\mbm{r}_i) \{ \mbm{\mu}_{n0} (\mbm{r}_j) \}^{\rm t}}{E_n -\hbar \omega -i \eta}
+ \frac{\mbm{\mu}_{n0} (\mbm{r}_i) \{ \mbm{\mu}_{0n} (\mbm{r}_j) \}^{\rm t}}{E_n +\hbar \omega +i \eta}
\right].
\end{equation}
Here, $\mbm{\mu}_{nm} (\mbm{r}_i) = -g\mu_{\rm B} \langle \mbm{k}_n|\hat{\mbm{S}}_i |\mbm{k}_m \rangle$.
The index $n$ represents the eigenstate $|\mbm{k}_n \rangle$ of $\hat{\mathcal{H}}_0$,
$\eta$ is an infinitesimal positive value due to causality, and
$v$ is the volume of the unit cell for each spin site.

We consider the magnetic nonlocal response of magnons in a thin ferromagnetic film.
At the surfaces, the spin waves are confined owing to the pinning effect.
For the magnon wavefunction, we obtain
$\sqrt{\frac{1}{v}} \langle 0| \hat{\mbm{S}}_i |{n,\mbm{k}_{\parallel}} \rangle
= \sqrt{\frac{S}{2}} \frac{1}{L} \exp (i\mbm{k}_{\parallel} \cdot \mbm{\rho}_{i,{\parallel}}) \phi_n (z_i)$,
where $\phi_n (z_i) = \sqrt{\frac{2}{d}} \sin (K_n z_i)$ and $K_n = n\pi/d$.
Here, $d$ and $L$ are the thickness of the film in the $z$-direction and the size of the system in the $xy$-plane, respectively.
$\mbm{k}_{\parallel}$ and $\mbm{\rho}_{i,{\parallel}}$ are the in-plane wavenumber and position vectors, respectively,
and $S$ is the value of the spin.
We apply the continuous approximation for the magnon wavefunctions.
For the translational symmetry in the $xy$-plane, the field and magnetization are partially Fourier-transformed as follows: 
$\mbm{h} (\mbm{\rho}_{\parallel},z;\omega) = (1/2\pi) \int {\rm d}^2 \mbm{k}_{\parallel} \ \tilde{\mbm{h}} (\mbm{k}_{\parallel},z;\omega)
e^{i\mbm{k}_{\parallel} \cdot \mbm{\rho}_{\parallel}}$.
The constitutive equation is also rewritten as
$\tilde{\mbm{m}} (\mbm{k}_{\parallel},z;\omega) = \int {\rm d} z' \bar{\chi} (\mbm{k}_{\parallel},z,z', \omega)
\tilde{\mbm{h}} (\mbm{k}_{\parallel},z';\omega)$, where the susceptibility is represented by
\begin{equation}
\bar{\chi} (\mbm{k}_{\parallel},z,z^\prime;\omega)
= -\frac{\mu_0}{v} \sum_n
\left[
\frac{\tilde{\mbm{\mu}}_n (z) \tilde{\mbm{\mu}}_n^\dagger (z^\prime)}{E_{n,\mbm{k}_{\parallel}} - \hbar \omega - i\eta}
+
\frac{\tilde{\mbm{\mu}}_n^* (z) \tilde{\mbm{\mu}}_n^{\rm t} (z^\prime)}{E_{n,\mbm{k}_{\parallel}} + \hbar \omega + i\eta}
\right].
\label{eq:singleX}
\end{equation}
where $\tilde{\mbm{\mu}}_n (z) = g\mu_{\rm B} \sqrt{\frac{S}{2}} \phi_n (z) (1,-i,0)^{\rm t}$.
For multi-layered systems, the susceptibility is expressed as a combination of Eq.\ (\ref{eq:singleX}) for the respective layers.

\subsection{Maxwell's equations}
\label{sec:Maxwell}
The coupling between the magnon and the electromagnetic field (photon) to form the MP is described by
a self-consistent relation with the nonlocal constitutive equation and the microscopic Maxwell's equations,
where the spatial correlation between the magnon wavefunction and the electromagnetic field are treated appropriately.
The microscopic Maxwell's equations are
$\mbm{\nabla} \cdot \mbm{E}  = -(1/\varepsilon_0) \mbm{\nabla} \cdot \mbm{P}$,
$\mbm{\nabla} \times \mbm{E} = -\partial \mbm{B} / \partial t$,
$\mbm{\nabla} \cdot \mbm{B}  = 0$,
and
$\mbm{\nabla} \times \mbm{B} = \varepsilon_0 \mu_0 \partial \mbm{E} / \partial t
+ \mu_0 ( \mbm{\nabla} \times \mbm{M} + \partial \mbm{P}/\partial t)$.
Here, $\mbm{E}$, $\mbm{B}$, $\mbm{M}$, $\mbm{P}$, $\varepsilon_0$, and $\mu_0$ denote 
the total electric field, total magnetic flux density, magnetization, polarization,
dielectric constant in vacuum, and permeability in vacuum, respectively.
The magnons are described in terms of $\mbm{M} (\mbm{r})$ with nonlocal susceptibility in Eq.\ (\ref{eq:singleX}),
while the polarization is represented by the dielectric constant of each layer,
$\mbm{P} (\mbm{r}) = \varepsilon_0 \chi_{\rm E} (z) \mbm{E} (\mbm{r})$.
The dielectric constant $\varepsilon_0 \chi_{\rm E} (z)$ forms the TE and TM waveguide modes.
We set the field propagation direction along the $xz$-plane, $\mbm{k} = (k_{\parallel},0,k_z)^{\rm t}$.
Subsequently, the $y$-component of the magnetic (electric) field determines
the reflection and transmission of the TM (TE) mode.

For the magnetic field $\mbm{H} = \mbm{B}/\mu_0 - \mbm{M}$, the Maxwell's equations are
$\mbm{\nabla} \times \mbm{\nabla} \times \mbm{H}
+ \mu_0 \varepsilon_0 (1 + \chi_{\rm E}) \partial^2 \mbm{H}/\partial t^2
= - \mu_0 \varepsilon_0 (1 + \chi_{\rm E}) \partial^2 \mbm{M}/\partial t^2
+ \varepsilon_0 (\mbm{\nabla} \chi_{\rm E}) \times \partial \mbm{E}/\partial t$.
The systems considered in this study exhibits transverse symmetry in the $xy$-direction.
We consider partial Fourier transformations with $\mbm{k}_{\parallel}$ and $\omega$ for the magnetic fields.
$\tilde{\mbm{H}} (\mbm{k}_{\parallel}, z;\omega) = \frac{1}{(2\pi)^{3/2}} \int {\rm d}x{\rm d}y \int {\rm d}t,
\mbm{H} (\mbm{r};t) e^{i(k_x x + k_y y)} e^{-i\omega t}$.
The electric field is also Fourier-transformed in the same manner.
For the TM mode, $\tilde{\mbm{H}}^{\rm (TM)} = \left( 0,\tilde{H}_y,0 \right)^{\rm t}$, the solution is given as
\begin{equation}
\tilde{\mbm{H}}^{\rm (TM)} (\mbm{k}_{\parallel}, z;\omega)
= \tilde{\mbm{H}}^{\rm (TM)}_{\rm inc} (\mbm{k}_{\parallel}, z ;\omega)
+ \int {\rm d}z^\prime \bar{G}^{\rm (TM)} (\mbm{k}_{\parallel}, z, z^\prime ;\omega)
\tilde{\mbm{M}} (\mbm{k}_{\parallel}, z^\prime ;\omega)
\label{eq:constitutiveTM}
\end{equation}
with a dyadic Green's function
\begin{equation}
\bar{G}^{\rm (TM)} (\mbm{k}_{\parallel} ,z ,z^\prime ;\omega)
= - \left( \begin{matrix}
0 &  &  \\
  & q(z^\prime)^2 &  \\
  &  & 0
\end{matrix} \right)
\frac{i g (\mbm{k}_{\parallel} ,z ,z^\prime ;\omega)}{2 \sqrt{q(z^\prime)^2 - k_{\parallel}^2}}.
\label{eq:GreenTM}
\end{equation}
Here, $g (\mbm{k}_{\parallel}, z , z^\prime ;\omega)$ is a scalar Green's function describing
a scalar wave propagation from position $z^\prime$ in a layer to $z$ in another layer.
The $(\mbm{\nabla} \chi_{\rm E}) \times \tilde{\mbm{E}}$ term is considered by
the boundary conditions for $g (\mbm{k}_{\parallel}, z , z^\prime ;\omega)$.
In the $n$-th layer, $z_{n-1} < z < z_n$, the wavenumber is $q (z) = q_n = \chi_{{\rm E},n} \omega/c$.
When $z_{n-1} < z < z_n$ and $z_{m-1} < z^\prime < z_m$, the scalar Green's function is given as
\begin{equation}
g (\mbm{k}_{\parallel},z,z';\omega)
= \left\{ \begin{matrix}
\left[ e^{ ik_n (z - z_{n-1})} + e^{-ik_n (z - z_n)} \tilde{r}_{n+1,n} e^{ ik_n d_n}     \right]
\tilde{t}_{n,m} A_m (z')
 & (n>m) \\
e^{ik_m |z - z'|} + e^{ik_m (z - z_{m-1})} B_m (z') + e^{-ik_m (z - z_m)} C_m (z')
 & (n=m) \\
\left[ e^{-ik_n (z - z_n)}     + e^{ ik_n (z - z_{n-1})} \tilde{r}_{n-1,n} e^{ ik_n d_n} \right]
\tilde{t}_{n,m} D_m (z')
 & (n<m)
\end{matrix} \right.
\label{eq:sGreen}
\end{equation}
where $\tilde{r}_{n \pm 1,n}$ and $\tilde{t}_{n,m}$ are the generalized reflection and transmission
coefficients, respectively~\cite{book:Chew}. Here,
$k_n = \sqrt{q_n^2 - k_{\parallel}^2}$ is the wavenumber in the perpendicular direction of layer $n$.
The factors are
\begin{eqnarray}
A_m (z') &=&
M_m \left[ e^{ ik_m (z_m - z')} + e^{ ik_m d_m} \tilde{r}_{m-1,m} e^{-ik_m (z_{m-1} - z')} \right],
\\
B_m (z') &=&
M_m \tilde{r}_{m-1,m} \left[ e^{-ik_m (z_{m-1} - z')} + e^{ik_m d_m} \tilde{r}_{m+1,m} e^{ ik_m (z_m - z')}    \right],
\\
C_m (z') &=&
M_m \tilde{r}_{m+1,m} \left[ e^{ ik_m (z_m - z')}     + e^{ik_m d_m} \tilde{r}_{m-1,m} e^{-ik_m (z_{m-1} - z')} \right],
\\
D_m (z') &=&
M_m \left[ e^{-ik_m (z_{m-1} - z')} + e^{ ik_m d_m} \tilde{r}_{m+1,m} e^{ ik_m (z_{m-1} - z')} \right],
\end{eqnarray}
and
$M_m = \left[ 1 - \tilde{r}_{m+1,m} \tilde{r}_{m-1,m} e^{i2k_m d_m} \right]^{-1}$.

For the electric field, we obtain
$\mbm{\nabla} \times \mbm{\nabla} \times \mbm{E}
+ \mu_0 \varepsilon_0 (1 + \chi_{\rm E}) \partial^2 \mbm{E}/\partial t^2
= - \mu_0 \mbm{\nabla} \times \partial \mbm{M}/\partial t$.
For the TE mode, $\tilde{\mbm{E}}^{\rm (TE)} = \left( 0,\tilde{E}_y,0 \right)^{\rm t}$,
is related with $\tilde{\mbm{H}}^{\rm (TE)}$ through Faraday's law.
The $\mbm{\nabla} \mbm{\nabla} \cdot \mbm{E}$ term from the left-hand side is incorporated with the boundary conditions.
Thus, the solution of
$\tilde{\mbm{H}}^{\rm (TE)} = \left( \tilde{H}_x,0, \tilde{H}_z \right)^{\rm t}$ is
\begin{equation}
\tilde{\mbm{H}}^{\rm (TE)} (\mbm{k}_{\parallel},z ;\omega)
= \tilde{\mbm{H}}^{\rm (TE)}_{\rm inc} (\mbm{k}_{\parallel},z ;\omega)
+ \int {\rm d}z' \bar{G}^{\rm (TE)} (\mbm{k}_{\parallel}, z, z^\prime;\omega)
\tilde{\mbm{M}} (\mbm{k}_{\parallel}, z^\prime;\omega)
\label{eq:constitutiveTE}
\end{equation}
with
\begin{equation}
\bar{G}^{\rm (TE)} (\mbm{k}_{\parallel}, z, z^\prime;\omega)
= - \left( \begin{matrix}
\partial_z \partial_{z^\prime} &  & ik_{\parallel} \partial_z \\
  & 0 &  \\
-ik_{\parallel} \partial_{z^\prime} &  & k_{\parallel}^2
\end{matrix} \right)
\frac{i g (\mbm{k}_{\parallel},z ,z^\prime ;\omega)}{2 \sqrt{q(z^\prime)^2 - k_{\parallel}^2} }
-
\left( \begin{matrix}
0 &   &  \\
  & 0 &  \\
  &   & 1
\end{matrix} \right)
\delta (z-z^\prime).
\label{eq:GreenTE}
\end{equation}
The scalar Green's function $g(\mbm{k}_{\parallel}, z, z^\prime ;\omega)$ is the same as that for the TM mode.
The $xx$-component of the second term is cancelled by the elimination of singularity owing to the $\partial_z \partial_{z^\prime}$ term.

Finally, we obtain the Green's function as the sum of Eqs.\ (\ref{eq:GreenTM}) and (\ref{eq:GreenTE}),
$\bar{G} (\mbm{k}_{\parallel}, z, z^\prime ;\omega) = \bar{G}^{\rm (TM)} (\mbm{k}_{\parallel}, z, z^\prime ;\omega)
+ \bar{G}^{\rm (TE)} (\mbm{k}_{\parallel}, z, z^\prime ;\omega)$.

\subsection{Self-consistent treatment}
\label{sec:selfconsistent}
The electromagnetic fields $\tilde{\mbm{H}}$ and $\tilde{\mbm{E}}$, the magnetization $\tilde{\mbm{M}}$, and
the polarization $\tilde{\mbm{P}}$ are determined self-consistently based on the constitutive and Maxwell's equations.
The self-consistent relation is rewritten in the matrix form as follows:
For the total and incident fields, the coefficients are introduced as
\begin{eqnarray}
F_n &=&
\int {\rm d}z \frac{\tilde{\mbm{\mu}}_n^\dagger (z) \tilde{\mbm{h}} (z)}
{E_{n,\mbm{k}_{\parallel}} - \hbar \omega - i\eta}, \\
f_n &=&
\int {\rm d}z \frac{\tilde{\mbm{\mu}}_n^{\rm t}   (z) \tilde{\mbm{h}} (z)}
{E_{n,\mbm{k}_{\parallel}} + \hbar \omega + i\eta}, \\
X_n &=&
\int {\rm d}z {\tilde{\mbm{\mu}}_n^\dagger(z) \tilde{\mbm{h}}_{\rm inc} (z)}, \\
x_n &=&
\int {\rm d}z {\tilde{\mbm{\mu}}_n^{\rm t}   (z) \tilde{\mbm{h}}_{\rm inc} (z)}.
\end{eqnarray}
The associated vectors are $\mbm{F} = ( \{ F_n\}, \{f_n\} )^{\rm t}$ and $\mbm{X} = ( \{ X_n\}, \{x_n\} )^{\rm t}$.
The dyadic Green's function yields the factors of the matrices:
\begin{eqnarray}
A_{n,n'} &=& -\frac{\mu_0}{v} \int {\rm d}z \int {\rm d}z'
\tilde{\mbm{\mu}}_n^\dagger (z) \bar{G} (\mbm{k}_{\parallel},z,z';\omega) \tilde{\mbm{\mu}}_{n'} (z'), \\
a_{n,n'} &=& -\frac{\mu_0}{v} \int {\rm d}z \int {\rm d}z'
\tilde{\mbm{\mu}}_n^{\rm t} (z) \bar{G} (\mbm{k}_{\parallel},z,z';\omega) \tilde{\mbm{\mu}}_{n'}^* (z'), \\
W_{n,n'} &=& -\frac{\mu_0}{v} \int {\rm d}z \int {\rm d}z'
\tilde{\mbm{\mu}}_n^\dagger (z) \bar{G} (\mbm{k}_{\parallel},z,z';\omega) \tilde{\mbm{\mu}}_{n'}^* (z'), \\
w_{n,n'} &=& -\frac{\mu_0}{v} \int {\rm d}z \int {\rm d}z'
\tilde{\mbm{\mu}}_n^{\rm t} (z) \bar{G} (\mbm{k}_{\parallel},z,z';\omega) \tilde{\mbm{\mu}}_{n'} (z').
\end{eqnarray}
We find
\begin{equation}
\bar{S} \mbm{F} = \mbm{X}
\label{eq:selfconsistent}
\end{equation}
with
\begin{equation}
\bar{S} \equiv \left( \begin{matrix}
(\bar{E} - \hbar \omega - i\eta) \bar{1} + \bar{A} & \bar{W}, \\
\bar{w} & (\bar{E} + \hbar \omega + i\eta) \bar{1} + \bar{a}
\end{matrix} \right).
\label{eq:Smatrix}
\end{equation}
$\bar{E}$ is a diagonal matrix based on eigenenergies.
Notably, the interaction between magnons is described in the matrix through the Green's function.
The determinant of matrix $\bar{S}$ yields the dispersion relation of the MPs.

Even for magnetic materials with magnon excitation,
the dielectric constant contributes to the microwave reflection and refraction.
The nonlocal self-consistent theory proposed herein for the magnons and the radiative field
plays a significant role, although we focus on the linear response of magnons.

\end{document}